\documentclass[aps,prd,twocolumn,superscriptaddress,floatfix,citeautoscript,hyperref=pdftex]{revtex4-1}
\usepackage[utf8]{inputenc}

\newcommand{\refb}[1]{(\ref{#1})}

\newcommand{\nn}{\nonumber}

\usepackage{color}
\usepackage[colorlinks=true,linkcolor=red,
            urlcolor=purple,
            citecolor=blue]{hyperref}
\usepackage{verbatim}
\usepackage{latexsym,amssymb,float}
\usepackage{setspace}
\usepackage{graphicx}
\usepackage{subfig}
\usepackage{epsfig}
\usepackage{amsmath}
\usepackage{bm}
\usepackage{lipsum}
\usepackage{tikz}
\usetikzlibrary{arrows,shapes}
\usetikzlibrary{trees}
\usetikzlibrary{matrix,arrows} 				
											
\usetikzlibrary{positioning}				
\usetikzlibrary{calc,through}				
\usetikzlibrary{decorations.pathreplacing}  

\usepackage{pgffor}							

\usetikzlibrary{decorations.pathmorphing}	
\usetikzlibrary{decorations.markings}
\tikzset{
    vector/.style={decorate, decoration={snake}, draw},
	provector/.style={decorate, decoration={snake,amplitude=2.5pt}, draw},
	antivector/.style={decorate, decoration={snake,amplitude=-2.5pt}, draw},
    fermion/.style={draw=black, postaction={decorate},
        decoration={markings,mark=at position .55 with {\arrow[draw=black]{>}}}},
    fermionbar/.style={draw=black, postaction={decorate},
        decoration={markings,mark=at position .55 with {\arrow[draw=black]{<}}}},
    fermionnoarrow/.style={draw=black},
    gluon/.style={decorate, draw=black,
        decoration={coil,amplitude=4pt, segment length=5pt}},
    scalar/.style={dashed,draw=black, postaction={decorate},
        decoration={markings,mark=at position .55 with {\arrow[draw=black]{>}}}},
    scalarbar/.style={dashed,draw=black, postaction={decorate},
        decoration={markings,mark=at position .55 with {\arrow[draw=black]{<}}}},
    scalarnoarrow/.style={dashed,draw=black},
    electron/.style={draw=black, postaction={decorate},
        decoration={markings,mark=at position .55 with {\arrow[draw=black]{>}}}},
	bigvector/.style={decorate, decoration={snake,amplitude=4pt}, draw},
}

\tikzstyle{block} = [draw, rectangle, 
    minimum height=3em, minimum width=6em]

\newcommand{\CE}{{\cal E}}

\newcommand{\be}{\begin{equation}}
\newcommand{\ee}{\end{equation}}

\begin{document}

\title{Supersymmetric black holes and $T\bar T$ deformation}

\author{Jan Manschot, Swapnamay Mondal
	\\
	 {\it School of Mathematics, Trinity College, Dublin 2, Ireland}
	\\
	 {\it Hamilton Mathematical Institute, Trinity College, Dublin 2, Ireland}
}
  
\begin{abstract}
The entropy of supersymmetric black 
holes in string theory compactifications can be related to that of a
D- or M-brane system, which in many cases can be further reduced to a two-dimensional conformal field
theory (2d CFT). For black holes in M-theory, this relation involves a decoupling limit where the black hole mass
diverges. We suggest
that moving away from this limit corresponds to a specific irrelevant
perturbation of the 2d CFT, namely the supersymmetric completion of the
$T\bar T$ deformation. We demonstrate that the black hole mass matches
precisely with the $T\bar T$ deformed energy levels, upon identifying the
$T\bar T$ deformation parameter with the inverse of the leading term
of the black hole mass. We discuss various implications of this novel
realization of the $T\bar T$ deformation, including a Hagedorn temperature for wrapped
M5-branes, and potential change of degeneracies in the deformed theory.
\end{abstract}
    

\maketitle

\par \noindent \textit{Introduction.---} \label{s1}
Black holes are known to host an enormous amount of entropy, known as
Bekenstein-Hawking entropy $S_{\rm BH}= \frac{A}{4\ell_P^2}$, where
$A$ is the area of the event horizon and $\ell_P$ is the Planck length
\cite{Hawking1, BCH, Bekenstein1, Bekenstein2, Hawking75}. This formula is
particularly striking since the black hole entropy scales as area
rather than volume. This intriguing observation has given rise to the
idea of holography \cite{tHooft:1993dmi, Susskind:1994vu}. Another
remarkable feature is the appearance of the Planck length, the
characteristic length scale of quantum gravity, although the entropy
formula was derived in the realm of classical gravity. This suggests
that understanding of black hole entropy is key to the understanding of quantum gravity.
The immediate puzzle is that black hole microstates can not
be distinguished and enumerated within classical gravity.

For supersymmetric black holes, one can overcome this stumbling
block, by utilising the fact that the degeneracy, or index to be
precise \cite{Witten_morse,Schellekens_Warner,Witten_elliptic}, 
is locally constant as function of continuous parameters of the theory. For black
holes in string theory, a decoupling limit in which Newton's
constant becomes small leads to a complimentary 
description in terms of D-branes, often described by a 2-dimensional conformal field theory (CFT). Microstates have a clear
meaning in this picture and their counting reproduces $S_{\rm BH}$
correctly \cite{Strominger:1996sh, Maldacena:1997de}.
It is an interesting question to extend the microscopic description
away from the limit.
  
In this article, we consider 4-dimensional compactifications of string theory with
$\mathcal{N}=2$ supersymmetry. This theory gives rise supersymmetric
black holes, that is to say half-Bogomol'nyi-Prasad-Sommerfield (BPS) states, since they preserve
four out of the eight supercharges.
There is a great
variety of these states, with intricate dependence on asymptotic
values of the vector multiplet scalars, which
parameterize the K\"ahler moduli space
\cite{Maldacena:1998uz, Maloney:1999dv, Denef:2000nb, Denef:2001xn, Ooguri:2004zv,
  Bena:2006kb, Denef:2007vg, deBoer:2008zn,
  Manschot:2010qz, Cheng:2007ch, Dabholkar:2012nd}. We are interested in a family of
black holes whose degrees of freedom reduce to a 2d CFT, more precisely the Maldacena-Strominger-Witten (MSW) CFT
\cite{Maldacena:1997de,Minasian:1999qn}, in a decoupling limit of the
K\"ahler moduli space. This limit corresponds to the ``infinite volume
limit'' in string units, $|J| \to \infty$ with $J$ being the real
K\"{a}hler modulus. Newton's
  constant $G_4$ vanishes in this limit.

For this family of black holes, moving to finite mass
coincides with moving away from the infinite volume limit. In this
article, we argue that moving to the ``finite volume regime" of a
family of half-BPS black
holes in $\mathcal{N}=2$ theory is captured by $T\bar T$ deformation
of the MSW CFT.  The $T\bar T$ deformation is an irrelevant
deformation constructed in terms of the energy-momentum tensor
\cite{Zamolodchikov:2004ce}. Even though irrelevant, the theory
remains solvable, and the degeneracies
do not lift under this deformation. Our main argument, that moving to
finite volume corresponds to the $T\bar T$ deformation is that the
expression for the black hole masses matches with the formula for
$T\bar T$ deformed energy levels.

\vspace{2mm} 
\par \noindent\textit{ D4-brane black holes \& CFT. ---}\label{s2}
Extremal black holes in $\mathcal{N}=2$ supergravity are characterised
by their magnetic and electric charges $\gamma=\left( P^0, P^a, Q_a,
  Q_0\right), \, a = 1,\ldots ,n_{v}$, were $n_v$ is the number of
vector multiplets. Such black holes preserve 4 out of total 8
supercharges.
By the attractor mechanism  for extremal black holes, irrespective of their values at spatial infinity, the scalars $t^a := \tfrac{Y^a}{Y^0}$ flow to their ``attractor values" $t^a_\gamma $ at the black hole horizon \cite{Ferrara:1995ih}. 
For the family of black holes with $P^0=0$, the Bekenstein-Hawking entropy of the
black hole is $S_{\rm BH} =\pi \sqrt{\frac{2}{3} P^3 \widehat{Q}_{\bar
    0}}$ \cite{Shmakova:1996nz}, where   
$P^3= d_{abc}P^{a} P^{b} P^{c}$, with $d_{abc}$ the 3-tensor of the tree-level
prepotential $F(Y) = \frac{d_{abc} Y^a Y^b Y^c}{6 Y^0}$, with $Y^a$-s
being the vector multiplet scalars.  Moreover, $\widehat{Q}_{\bar 0}= -Q_{0} +
\dfrac{1}{2}d^{ab}Q_{a} Q_{b}$, with $d^{ab}$, being the inverse of
the quadratic form $d_{ab}=d_{abc}P^{c}$. The signature of $d_{ab}$ is $(1,b_2-1)$.
 
$\mathcal{N}=2$ supergravity can in turn be realized as low energy
effective description of string theory, compactified on a Calabi-Yau
3-fold $X$. In type IIA string theory description, the charge $\gamma$ is carried by D6-D4-D2-D0 branes, wrapped around appropriate cycles of $X$. Number of vector multiplets $n_v$ equals the second Betti number $b_2$ of $X$. 
In type IIB string theory, these states manifest themselves in hypermultiplet geometry \cite{Alexandrov:2012au, Alexandrov:2013yva, Alexandrov:2016tnf}.

In the large volume limit, the Arnowitt-Deser-Misner (ADM) mass
$M_\gamma(t)$ of a D4-brane black hole carrying charge $\gamma= (0,P, Q, Q_0)$ equals
$|Z(\gamma,t)|$ with $Z(\gamma,t)$ the tree level holomorphic central charge, 
$Z(\gamma,t)=-\frac{1}{2 \ell_s^{5}}Pt^2+ \frac{1}{\ell_s^3}Qt-\frac{Q_0}{\ell_s}$, where $\ell_s$ is the string length, 
 $t^a= \frac{Y^a}{Y^0} = B^a + i J^a$ are the complexified K\"{a}hler
 moduli fields. We will also use $t^a=B^a+i\lambda \underline J^a$, with
 $\underline J$ the normalized K\"ahler modulus,  $P\underline
 J^2=\ell_s^4$. Thus $\lambda\sim (V_X/\ell_s^{6})^{1/3}$ is a dimensionless measure for the
 volume of 2-cycles in string units. The infinite volume limit then corresponds to $\lambda\to \infty$.
We will set $\ell_s=1$, unless otherwise
 mentioned. The (renormalized) rest mass $M_\gamma(t)$  simplifies in the infinite volume limit
 $\lambda\to \infty$\cite{Denef:2007vg, Andriyash:2008it, Manschot:2009ia, Manschot:2010xp, deBoer:2008fk}, rendering various supergravity partition functions amenable to analytics. Note the black hole mass $M_\gamma(t)$ diverges in this limit.

For vanishing D6-brane charge, type IIA string theory can be further
uplifted to M-theory, by introducing the M-theory circle  $S^1_M$ with radius $R = g_s \ell_s/2\pi$, where $g_s$ is the string coupling. In M-theory the D4-brane is elevated to a M5-brane wrapping the 4-cycle $P$ times the M-theory circle $S^1_M$. D2-brane charges are realised as fluxes in the world volume of the M5-brane, whereas the D0-charge becomes the momentum along $S^1_M$.
The M5-brane dynamics is captured by MSW CFT in the decoupling limit
\cite{Maldacena:1997de,Minasian:1999qn}. In this limit,  the
11-dimensional Planck length vanishes, $\ell_{11}=g_s^{1/3}\,\ell_s\to
0$, such that
\begin{align} 
R/\ell_{11} \rightarrow \infty, \quad V_X/\ell_{11}^6 \text{ fixed but large} \, . \label{decouple}
\end{align}
The second quantity is fixed since it corresponds to a hypermultiplet scalar. 

The MSW CFT has $(4,0)$ supersymmetry, that is to say 4 chiral
supercharges, which matches the number of preserved supersymmetries of
the black hole. The bosons of the theory comprise 3
non-chiral scalars corresponding to movement of the brane system along
transverse $\mathbb{R}^3$. In addition, there are non-chiral real scalars describing movement of the D4-brane inside
$X$, as well as left-moving
and right-moving scalars
descending from the M5-brane worldvolume self-dual 3-form field strength. 
The bosons and their fermionic partners can be arranged in $(4,0)$
supermultiplets, whose numbers depend on Calabi-Yau data as well as
the divisor. The four supersymmetries broken by the brane
configuration lead to a universal Goldstino supermultiplet
\cite{Minasian:1999qn}. The left/right central charges arising from
field content are $c_L = P^3 + \frac{1}{2} c_2.P$ and $c_R = P^3 +
c_2.P$, with $c_2$ the second Chern class of $X$.
The combination $\widehat{Q}_{\bar
  0}$ is bounded below by $-c_L/24$ with $c_L$ the central charge of
the left-moving degrees of freedom.
The Cardy formula for the left-movers reproduces the 1-loop corrected Bekenstein-Hawking entropy\cite{Maldacena:1997de}. 

On the gravity side, D4-brane black holes develop AdS$_3$ throats
\cite{deBoer:2008fk} dual to the MSW CFT, after uplifting from 4 to 5
dimensions \cite{Gaiotto:2005gf, Gaiotto:2005xt} and then taking the decoupling limit \refb{decouple}.
Since $\lambda\sim V_X^{1/3}/\ell_s^2=(V_X/\ell_{11}^6)^{1/3}
(R/\ell_{11})$, the limit (\ref{decouple}) is equivalent to the infinite volume limit in string units, $\lambda\to
\infty$.

An important aspect which complicates the state counting of single
center black holes are multicenter black holes \cite{Denef:2000nb,
  Denef:2007vg, Denef:2002ru, Bates:2003vx}. Their low energy dynamics
is captured by $\mathcal{N}=4$ quiver quantum mechanics
\cite{Denef:2002ru}. Five dimensional multicenter solutions with
centers carrying non-vanishing D6-brane charge have a distance scale
$\sim \ell_5^3/R^2\sim R/\lambda^3$, whereas those with vanishing
D6-brane charge have a distance scale $\sim \ell_5$, with
$\ell_5=\ell_{11}^3/(4\pi V_X^{1/3})$ the 5-dimensional Planck length.
Upon appropriate coordinate redefinition, configurations with distance
scale $\sim \ell_5^3/R^2$ go over to a single $AdS_3$ throat in this limit. On the other hand,
multicenter solutions with distance scale $\sim \ell_5$ form multiple throats  \cite{deBoer:2008fk}. 
Scaling black holes \cite{Denef:2007vg, Bena:2012hf,
  Chattopadhyaya:2021rdi, Beaujard:2021fsk} with centers carrying
vanishing D6-brane charges present an intriguing case, since these can
approach each other arbitrarily close. In 4 dimensions, the near coincident regime of scaling black holes develop an AdS$_2$ factor \cite{Anninos:2013nra, Mirfendereski:2020rrk}, thus in five dimensions they are expected to form an AdS$_3$ throat. 
The black holes to go over to a single AdS$_3$ throat are expected to
be captured by the CFT. In terms of the K\"ahler moduli, the 4d supergravity states which correspond to the CFT are those which
exist at the infinite volume attractor point \cite{deBoer:2008fk, Alexandrov:2016tnf}, 
$t^\infty_\gamma=\lim_{\lambda\to \infty} t^\lambda_\gamma$, with
\be
\label{tlg}
(t^\lambda_\gamma)^a=d^{ab}Q_b\,\ell_s^2+i \lambda\, \ell_s^2\, \frac{P^a}{\sqrt{P^3}}.
\ee
Then $\lambda=(p^3)^{1/6}(6^{1/3}/2)\,R/\ell_5$. Note that the
attractor value $t^\lambda_\gamma$ (\ref{tlg}) differs for different
D2-brane charge, even for states within the same CFT.

The ground state of the CFT corresponds to the ``bare'' D4-brane, whereas
excited states carry additional D0- and D2-brane charge. The energies
of excitations of the CFT correspond to the infinite volume limit of the D4-brane
mass, renormalized by subtracting the leading term $PJ^2/2$. For
$B=0$, this gives for the CFT energy $E_\gamma$ and momentum $\Pi_\gamma$,
\begin{equation}
\begin{split}
  RE_\gamma&= \lim_{|J|\to \infty} \ell_s\left(M_\gamma(t) -
  \tfrac{1}{2}PJ^2/\ell_s^5\right)\\
&= -Q_0+\frac{(Q.J)^2}{PJ^2},\\
R\Pi_\gamma&= Q_0 \, . 
\end{split}
\label{EPi}
\end{equation}
The expression for $B \neq 0$ is invariant under translations in the
electric-magnetic duality group Sp$(2+2b_2,\mathbb{Z})$.
Note $Q_0$ is the momentum along the M-theory circle $S^1_M$.
Eq. \refb{EPi} in turn implies for the Virasoro operators, $L_0 =
\frac{(Q.J)^2}{2PJ^2} + \frac{c_L}{24}$ and $\bar{L}_0 = -Q_0+\frac{(Q.J)^2}{2 PJ^2} + \frac{c_R}{24}$.
$L_0$ saturates the BPS bound for half-BPS states, which preserve 4 fermionic symmetries.

Altogether, one has the CFT partition function \cite{deBoer:2006vg,
  Gaiotto:2006wm, Denef:2007vg},
\begin{equation}
  \label{ZCFT}
\begin{split}
\mathcal{Z}_{\rm CFT}(\tau,\bar \tau) &= \sum_{Q_0, Q} \Omega(\gamma;t^\infty_\gamma)\, q^{(E_\gamma + \Pi_\gamma)/2}\, \bar{q}^{(E_\gamma - \Pi_\gamma)/2} \\
&\quad \times \int \, d^3\vec p \, e^{-\beta \frac{\vec p^2}{2 m_5}} \, ,
\end{split}
\end{equation}
where $m_5 = \frac{\pi PJ^2}{g_s \ell_s^5}$ is the mass of the wrapped
MSW string, and $\vec p$ is the momenta in $\mathbb{R}^3$. The
coefficient $\Omega(\gamma;t^\infty_\gamma)$ is the appropriate (rational) BPS index
\cite{Manschot:2010qz, Alexandrov:2016tnf}; it is a specialization of
$\Omega(\gamma;t)$ which is independent of
hypermultiplet scalars, while only locally constant as
function of the vector multiplet scalars through itos dependence on $t$.
Moreover, $q=e^{2\pi i \tau}$ with $\tau = C_0 + i \frac{\beta}{\ell_s g_s}$ 
the modular parameter of the torus $S^1_M \times S^1_\beta$, where
$S^1_\beta$ is the thermal circle, $\tau_2 := \text{Im} (\tau)$ is the ratio of circumferences
of $S^1_\beta$ and $S^1_M$. Similarly, $\tau_1 := \text{Re}(\tau)$ is
related to the IIA Ramond-Ramond 1-form $C_1$ as $C_1 = C_0
\frac{dt}{\beta}$ and describes the tilt of $S^1_\beta$ with respect to $S^1_M$.
We suppress further
non-holomorphic contributions related to mock modular forms
\cite{Manschot:2009ia, Alexandrov:2016tnf, Alexandrov:2018lgp}, which
are not relevant for the present discussion.

\vspace{2mm}  
\par\noindent \textit{Modularity.---}\label{secmod}
The MSW CFT is invariant under large reparameterizations of the torus,
that is to say modular transformations:
$\tau \rightarrow \frac{a\tau + b}{c \tau + d}, \, \, a,b,c,d \in \mathbb{Z}, \, \, ad-bc=1.$
The modular weight of $\mathcal{Z}_{\rm CFT}(\tau,\bar
\tau)$ is (2,0), since $\Omega(\gamma;t)$ is the second helicity
supertrace. The integral on the second line in Eq. (\ref{ZCFT}) is the integral over the three
momenta, which evaluates to a factor $\left( \frac{2\pi^2 P J^2}{g_s^2
    \tau_2} \right)^{3/2}=\left( \frac{2\pi^2 P J^2}{\beta g_s} \right)^{3/2}$.  This factor has modular weight 
$(\tfrac{3}{2}, \tfrac{3}{2})$, since $J$ has weight
$(\tfrac{1}{2},\tfrac{1}{2})$ and the area
of the torus $\beta g_s$ is modular invariant. 
The spectrum allows for a theta function
decomposition of $\mathcal{Z}_{\rm CFT}$, such that the degeneracies are enumerated by a weakly holomorphic
vector-valued modular form of weight $-1-b_2/2$ \cite{deBoer:2006vg,
  Gaiotto:2006wm, Denef:2007vg}. These functions can be explicitly
determined in special cases \cite{Gaiotto:2006wm, Gaiotto:2007cd,
  Collinucci:2008ht, VanHerck:2009ww, Alexandrov:2022pgd}
 
Compactifying M-theory on $S^1_M$ and further applying T-duality along
the $S^1_\beta$, one arrives at type IIB string theory. This theory exhibits an
$SL(2, \mathbb{Z})$ $S$-duality group. In this context, the modularity of the CFT
partition function and $S$-duality reinforce each other in a
non-trivial way \cite{Manschot:2009ia, Alexandrov:2012au, Alexandrov:2016tnf}.

\vspace{2mm} 
\par\noindent \textit{D4-brane black holes at finite volume.---} \label{s3}
Let us now turn to black holes with finite $\lambda$.  The mass
$M_\gamma(t)$ can be expressed in terms of $E_\gamma$
and $\Pi_\gamma$ (\ref{EPi}) as
\be
\label{Mgamma}
M_\gamma(t)=\frac{1}{\ell_s}\sqrt{(\tfrac{1}{2}PJ^2/\ell_s^4)^2+ (PJ^2/\ell_s^4)\,RE_\gamma+
  R^2\Pi_\gamma^2}\, .
\ee

Next, we introduce the energy  $\CE_\gamma(t)$ through, 
\be
\label{cegamma} 
R\CE_\gamma(t) = \ell_s M_\gamma(t) -  \tfrac{1}{2}PJ^2/\ell_s^4
\ee
for finite $PJ^2$. In the infinite volume limit, $\CE_\gamma$ simply reduces to $E_\gamma$ \refb{EPi}, the energy levels of the MSW CFT. Thus for finite $J$, we expect $\CE_\gamma$ to correspond to the energy spectra of the microscopic theory describing attractor black holes in finite volume regime. 

Remarkably with \refb{Mgamma} substituted, $\mathcal{E}_\gamma(t)$ (\ref{cegamma}) is
precisely of the form of the energy levels of the $T\bar{T}$
deformation of a two-dimensional CFT. This motivates us to propose that at finite volume, the microscopic description of D4-brane black holes is furnished by $T\bar{T}$ deformation of MSW CFT. 
We provide further justifications for the proposal in remainder of this article. 

\vspace{2mm} 
\par \noindent \textit{$T\bar T$ deformation.---} \label{s4} 
In a seminal paper \cite{Zamolodchikov:2004ce}, Zamolodchikov showed
that for a  two-dimensional quantum field theory, the composite
operator $T\bar{T}$ is free from short distance divergences, even
though this operator is an irrelevant operator modifying the ultraviolet behaviour of the theory.
The action of $S^\mu$ of the $T\bar T$ deformation of conformal field theory,
satisfies $dS^{(\mu)}/d\mu=\int dz\,d\bar z\, T^{(\mu)}\bar T^{(\mu)}$, with the stress energy
tensors those of the deformed theory with action
$S^{(\mu)}$. Remarkably, the energy levels $ E_n(R, \mu)$
  of the deformed theory can be determined exactly
  \cite{Smirnov:2016lqw, Cavaglia:2016oda} in terms of the momenta
  $P_n(R)$ and undeformed energy levels $E_n(R, 0)$ 
\begin{equation} 
  \label{edeform}
  \begin{split}
&  E_n(R, \mu) = - \frac{R}{2\mu} \\
  & \quad             +  \left[ \frac{R^2}{4\mu^2} + \frac{R}{\mu} E_n(R,0) +
                  P_n(R)^2 \right]^{1/2},
                \end{split}
              \end{equation}
where $R$ is the radius of the compact spatial dimension.
Recent years have seen a flurry of activities in this topic
\cite{Giveon:2017nie,Chakraborty:2020swe,Apolo:2019zai,Hashimoto:2019hqo,Hashimoto:2019wct,Chakraborty:2019mdf,Giveon:2019fgr,Chakraborty:2018vja,Callebaut:2019omt,Guica:2019vnb,Li:2020pwa,He:2020udl,He:2019ahx,
  Cardy:2019qao, Giribet:2017imm, Baggio:2018rpv, Ebert:2020tuy,
  Chang:2019kiu, Jiang:2019hux, Chang:2018dge, Shyam:2017znq,
  Asrat:2017tzd, Giveon:2017myj, Dubovsky:2017cnj, McGough:2016lol,
  Chakraborty:2020nme, Li:2020zjb, Brennan:2020dkw, Gross:2019ach,
  Kraus:2018xrn, He:2020hhm, Coleman:2020jte, Hartman:2018tkw,
  Cardy:2018sdv, Guica:2017lia,
  Bzowski:2018pcy,Aharony:2018ics,Datta:2018thy,
  Taylor:2018xcy,Rosso:2020wir,Dubovsky:2018bmo, Ferko:2021loo}. 
The $T\bar T$ deformation may change the UV dynamics of the theory. This is reflected in high energy density of states, which exhibits Hagedorn growth with Hagedorn temperature $T_H \sim 1/\sqrt{\mu}$, as opposed to Cardy growth.

The deformed energy levels satisfy an inviscid Burgers equation, which
exhibits shock singularities. This is related to the singularity which
arises for $E_n(R,0)<0$, that is to say the low lying states in a CFT.
For such states, $E_n(R,\mu)$ becomes imaginary for sufficiently large
$\mu$. At the crossover, the expression under the square-root in Eq. \refb{edeform} vanishes. 

Now let us consider the $T\bar T$ deformation of the MSW CFT. Then
Eq. (\ref{edeform}) demonstrates that the deformed energy levels equal
$\CE_\gamma (t)$, with the identification
\be
\label{RmuPJ}
R^2/\mu \longleftrightarrow PJ^2/\ell_s^4.
\ee
This suggests that at finite $\ell_{11}$, the M5-brane degrees of
freedom correspond to a $T\bar T$ deformed CFT. The identification
demonstrates that $\mu$ scales as $\ell_5^2$. Moving away from
 the infinite volume attractor point (\ref{tlg}) is naturally an
 irrelevant deformation. This can be analyzed in detail for the D1-D5 system with
 $(4,4)$ supersymmetry in two dimensions \cite{deBoer:2008ss}.

The shock waves mentioned above are in the black hole context related
to wall-crossing phenomena. The crossover happens when $M_\gamma(t)$ vanishes,
i.e. BPS states become massless. This situation can only arise for polar states, i.e. states with $\widehat
Q_{\bar 0} <0$. These states are realized as $D6$-$\overline{D6}$ bound states and they decay across a wall of
marginal stability, before reaching the massless point \cite{Denef:2007vg}.

For various free theories, even the action of the deformed theory can
be determined exactly \cite{Cavaglia:2016oda, Bonelli:2018kik} and has
the form of non-local Dirac-Born-Infeld (DBI) action, which arises as
effective action for D-branes. Although D-branes spontaneously break
some of the supersymmetry, the DBI action realizes the full Poincar\'{e}
symmetry, albeit in a non-linear fashion.

\vspace{2mm} 
\par\noindent\textit{Supersymmetry.---}\label{secSUSY}
Irrespective of the moduli, half-BPS black holes in $\mathcal{N}=2$
string theory preserve four supercharges, which must also be the case
for their microscopic description. This works out nicely in the
infinite volume limit, since the MSW
CFT has $(4,0)$ supersymmetry. As per our proposal, this requires for finite volume
regime  $(4,0)$ supersymmetry of the  $T\bar{T}$
deformed theory.  We expect that such a supersymmetric completion of
the $T\bar{T}$ deformation can be derived, as was obtained earlier for 
theories with (1,1) and (1,0) supersymmetries,
\cite{Chang:2018dge} and (2,2) and (0,2) supersymmetries \cite{Chang:2019kiu, Jiang:2019hux}. The
supersymmetric completion of the $T\bar T$ deformation for the MSW CFT
will be left invariant by the SO(4) R-symmetry. 

Along with four preserved supercharges, there are four broken supersymmetries leading to Goldstinos, which realize supersymmetry in a non-linear fashion. This is evident in MSW CFT \cite{Maldacena:1997de} and is expected to continue after $T\bar{T}$ deformation. It is encouraging that for free supersymmetric seed theory, the $T\bar{T}$ deformed theory is known to realize supersymmetry in a non-linear fashion \cite{Cribiori:2019xzp, Ferko:2019oyv}.

\vspace{2mm} 
\par\noindent\textit{Including momenta.---}  
Until now we have worked in the rest frame of the black hole. To
include the space-time momenta in the deformed theory, recall that the
rest mass $\frac{2\pi}{g_s}M_\gamma$ (as in the conventions of
\cite{Denef:2007vg}) is replaced by $\sqrt{ \left( \frac{2\pi}{g_s}
    |Z| \right)^2 + \vec{p}^2}$ in a moving frame. Eq. \refb{cegamma}
is then generalized to  
\begin{equation}
\begin{split}
  \mathcal{H}_\gamma(t)&= - \frac{1}{2} PJ^2\\
&+\sqrt{ (\tfrac{1}{2}PJ^2)^2+ PJ^2 \, E_\gamma  + \Pi_\gamma^2 + \left(\frac{g_s}{2\pi} \right)^2 \vec{p}^2},
\end{split}
\end{equation}
This reduces in the large volume limit to $E_\gamma+\vec p^2/2m_5$ (\ref{ZCFT}).

\vspace{2mm} 
\par\noindent\textit{Modularity revisited.---}  
It has been established that the $T \bar T$ deformation preserves
modularity of the partition function
\cite{Datta:2018thy,Aharony:2018bad}, which has modular weight (0,0) if the
dimensionless deformation parameter $\mu/R^2$ has weight $(-1,-1)$,
\be
\frac{\mu}{R^2}\to \frac{1}{|c\tau+d|^2} \frac{\mu}{R^2}.
\ee 
Clearly, the identification of the deformation parameter (\ref{RmuPJ}) 
in our model agrees with this transformation.
We can thus expect that the partition function of the modified MSW CFT
is also modular invariant. On the other hand, since the weight of the elliptic genus is
non-trivial, the relation between the deformed and undeformed elliptic
genus is more complicated \cite{Cardy:2022mhn}. The Type IIB
perspective \cite{Alexandrov:2012au, Alexandrov:2013yva} can be
relevant for this question, since it is valid for finite BPS mass.

\vspace{2mm} 

\par\noindent\textit{Holography.---}
An important question is the holographic dual of the $T\bar T$
deformed CFT, which has been addressed in various papers
\cite{McGough:2016lol, Giveon:2017myj, Giribet:2017imm, Asrat:2017tzd,
  Kraus:2018xrn, Guica:2019nzm, Chakraborty:2020swe}. As a first step
towards holography for the model discussed above, we present the metric
of the 5-dimensional uplift of a single centered black hole at the
large volume attactor point (\ref{tlg}) and including the dependence
on $\ell_5$, 
\begin{align}
\nn 
  \frac{1}{\ell_5^2} \, ds_{5d}^2
&=
\mathcal{N}^{-1}  \left[ - \frac{( \rho^2 - \rho_*^2)^2 }{\rho^2} dt^2 +  \rho^2 \left( d\alpha +  \frac{\rho_*^2}{\rho^2}  dt \right)^2 \right] \\
&+  \mathcal{N}^2  \left[ \frac{ 4 U^2 \rho^2 d\rho^2 }{(\rho^2 - \rho_*^2)^2 } + U^2 d\Omega_2^2 \right], \label{metric}
\end{align}
 with $\mathcal{N}  =  1 + \ell_5^2 (\rho^2 - \rho_*^2)  \, , \,
 \rho_*^2 := - \frac{4 \widehat{Q}_0}{U R^2}, \, U^3= P^3/6$.
 The 4d radial coordinate $r$ is in terms of these variables
 $r=\ell_5^3U(\rho^2-\rho_*^2))$.  
We have seen above that $\ell_5^2$ scales as
$T\bar T$ deformation $\mu$ in the decoupling limit (\ref{decouple}).
Indeed in this limit, the $(t,\alpha, \rho)$ coordinates of the metric
(\ref{metric}) parametrise a BTZ black hole \cite[Eq. (4.4)]{deBoer:2008fk}.

As per UV/IR correspondence in holography \cite{Susskind:1998dq}, the
bulk asymptotic region and deep interior are related to
ultraviolet and infrared respectively in the boundary theory. Since the $T \bar T$ deformation is irrelevant,
we expect the bulk asymptotics of (\ref{metric}) to change, yet the metric near the BTZ
singularity to remain unchanged. Indeed, the $\ell_5^2$ deformation
becomes negligible near the BTZ singularity for the ``infrared" limit 
$\rho\to \rho_*$, whereas in the ``ultraviolet"/asymptotic region $\rho \rightarrow \infty$ the geometry
asymptotes to $\mathbb{R}^2\times S^1$.

We naturally expect that the holographic description of the $T\bar T$ deformed model
is in terms of metrics which have the same asymptotics as (\ref{metric}) for $\rho\to \infty$.
An important complication for this analysis stems from the fact that
even small $\ell_5$ effects can not be treated as perturbation.
Eg.
if \refb{metric} is expanded to $\mathcal{O}(\ell_5^2)$, then the
resultant metric has wrong signature in the  asymptotic region $\rho^2
> \ell_5^{-2} + \rho_*^2$. A resolution may be to put a cut-off, as
proposed in \cite{McGough:2016lol, Kraus:2018xrn}, but here for the opposite
sign of  the $T \bar T$ deformation. 
In fact, the closest analog of asymptotic $AdS_3$, is the crossover
region $\rho_*^2 \ll \rho^2 \ll 1/\ell_5^2$, which exists whenever
$\ell_5^2 \ll 1/\rho_*^2$ and might be important for holography for
$\ell_5 \neq 0$. We leave further study for future work.

\vspace{2mm}  
  
\par\noindent\textit{Hagedorn transition.---}
Keeping with Cardy formula \cite{Cardy}, the degeneracies
$\Omega(\gamma; t_\gamma^\infty)$ grow as exponential of
$\pi\sqrt{\frac{2}{3}P^3\widehat Q_{\bar 0}}$ for large $\widehat
Q_{\bar 0}$. In general, $E_\gamma$ is bounded below by $\widehat
Q_{\bar 0}$, which is positive in the Cardy regime, while in the limit of
large D2-brane charge with $Q_0$ fixed, $2\widehat
Q_{\bar 0} \leq E_\gamma$. In the latter regime, the energy $\mathcal{E}_\gamma$ behaves
as $\sqrt{PJ^2\,E_\gamma}$. With the lower bound for $E_\gamma$, we
then have
\be
\label{HagTemp}  
\Omega(\gamma; t_\gamma^\infty)\,e^{-\mathcal{E}_\gamma/T}\leq e^{\pi
  \sqrt{\frac{2}{3}P^3\,\widehat Q_{\bar 0}}-\sqrt{2 PJ^2\,\widehat
    Q_{\bar 0}}/T} \, .
\ee
Consequently, the system gives rise to a Hagedorn temperature $T_H$
above which the sum over D2-charges diverges. Eq. (\ref{HagTemp})
shows that $\frac{1}{\pi}\sqrt{3\,PJ^2/P^3}$ is an upper bound for $T_H$.

The Hagedorn temperature \cite{Hagedorn} indicates the existence of a different high temperature phase. 
Various systems in string theory, including Little String Theory \cite{Kutasov:2000jp}, superstring theory  \cite{Bala, Kogan, Atwit}, $\mathcal{N}=4$ super Yang-Mills on compact spaces \cite{Sundborg:1999ue, Witten:1998zw}, BFSS matrix model in the presence of an IR cutoff \cite{Sathiapalan:2008ye} exhibit Hagedorn transition.
In the present case, since the relevant two-dimensional theories
descend from M5-brane world volume theory, this predicts a Hagedorn
temperature for wrapped M5 branes.

\vspace{2mm} 

\par\textit{Change of degeneracies?---} It has been argued that the
degeneracies are not lifted under the
$T\bar T$ deformation \cite{Zamolodchikov:2004ce}, and similarly that
supersymmetric indices remain unchanged \cite{Ebert:2020tuy}. This
suggests that the $\Omega(\gamma;t^\infty_\gamma)$ remain the
same as function of $\gamma$. On the other hand, moving to finite volume suggests that the
natural BPS index is $\Omega(\gamma;t^\lambda_\gamma)$ for finite
$\lambda$. While for simple systems of D4-branes, such as those with a
irreducible magnetic charge, these indices are indeed equal, this may
not be the case for more involved systems. For example,
Ref. \cite{Chattopadhyaya:2021rdi} described a family of scaling solutions, whose
degeneracy is subleading to that of a single center black hole. 
These solutions are present for finite $\lambda$, but decouple in the limit
$\lambda\to \infty$ \cite{deBoer:2008fk}. Thus the supergravity picture gives some
indication that degeneracies may change upon the $T\bar T$ 
deformation in sufficiently intricate CFT's. Of course for
sufficiently large $\mu$, the deformation may give rise to wall-crossing, under which
the degeneracies will also change.


\vspace{2mm} 
\par\noindent\textit{Discussion.---}  
We have presented evidence that the microscopic description
of D4-brane black holes with finite mass is furnished by a $T\bar{T}$
deformation of MSW CFT.

A first principle derivation of $T \bar T$
  deformation from the microscopic side is clearly desirable. One possibility is that this
  arises from integrating out gravitational effects. This could also
  explain non-locality of the $T\bar T$ deformed theory, and is, in fact, the case for infinitesimal $T \bar T$ deformations \cite{Cardy:2018sdv}.

Lastly, it would be interesting to explore whether a similar picture
extends to other microscopic descriptions of black holes such as the
D1-D5 system. 

\vspace{1mm}  

{\it Acknowledgement:} 
This work was supported by Laureate Award 15175 ``Modularity in
Quantum Field Theory and Gravity" of the Irish Research Council. We
thank Sergey Frolov, Nissan Itzhaki, Greg Moore and Herman Verlinde for correspondence
and discussions.
 
%

\end{document}